\documentclass{aa}
\input psfig.tex
\usepackage{graphicx}
\usepackage{natbib}

\usepackage{array}
\usepackage{graphics}
\usepackage{latexsym}
\usepackage{amssymb}
\usepackage{amsmath}
\usepackage{fancyhdr}
\usepackage{morefloats}
\usepackage{bm}
\bibpunct{(}{)}{;}{a}{}{,}

\begin{document}
\title{Orbital evolution of the Carina dwarf galaxy and self-consistent star formation history determination}

\author{S. Pasetto$^{1,2}$, E.K. Grebel$^{1}$, P. Berczik$^{1,4,5}$, C. Chiosi$^{3}$, R. Spurzem$^{1,4}$}

\institute{$^1$ Astronomisches Rechen-Institut, Zentrum f\"ur Astronomie der Universit\"at Heidelberg, Germany \\
$^2$ Max-Planck-Institut f\"ur Astronomie, Heidelberg, Germany\\
$^3$ Department of Astronomy, University of Padova, Italy\\
$^4$ National Astronomical Observatories of China (NAOC), Chinese Academy of Sciences (CAS), Datun Lu 20A, Chaoyang District, Beijing 100012, China\\
$^5$ Main Astronomical Observatory (MAO), National Academy of Sciences of Ukraine (NASU), Akademika Zabolotnoho 27, 03680 Kyiv, Ukraine\\
\email{{spasetto\char64ari.uni-heidelberg.de}}}
\date{Received: ;  Revised: ; Accepted 08-Sept-2010 in A\&A}

\titlerunning{Carina dwarf galaxy}
\authorrunning{S. Pasetto et al.}

\abstract{We present a new study of the evolution of the Carina dwarf galaxy that includes a simultaneous derivation of its orbit and star formation history. 
The structure of the galaxy is constrained through orbital parameters derived from the observed distance, proper motions, radial velocity and star formation history.
The different orbits admitted by the large proper motion errors are investigated in relation to the tidal force exerted by an external potential representing the Milky Way (MW). 
Our analysis is performed with the aid of fully consistent N-body simulations that are able to follow the dynamics and the stellar evolution of the dwarf system in order to determine self-consistently the star formation history of Carina. We find a star formation history characterized by several bursts, partially matching the observational expectation. We find also compatible results between dynamical projected quantities and the observational constraints. The possibility of a past interaction between Carina and the Magellanic Clouds is also separately considered and deemed unlikely.
 }

\maketitle

\section{Introduction}\label{Introduction}
It has been extensively reviewed by many authors (e.g., \citet{1999IAUS..192...17G, 1999A&ARv...9..273V, 1998ARA&A..36..435M}) that dwarf spheroidal galaxies (dSphs) are the most common type of galaxies in the Local Group (LG). Their origin and evolution are focal points because of the key role that they play in the context of the hierarchical growth of structures in the Universe, and because of the natural interest that these peculiar systems engender as examples of dark matter dominated systems. In the context of dynamical research on interacting systems, dSphs also play a crucial role as stellar systems found in the central regions of galaxy clusters and groups, thus becoming potential subjects of strong gravitational interactions. This morphology-position relation is also observed in our LG  \citep[e.g.,][]{1999IAUS..192...17G}.

Tidal interactions may leave a variety of structural signatures (see, for instance, \citet{2006MNRAS.367..387R,2008ApJ...679..346M,2008ApJ...673..226P,2009ApJ...698..222P}). In the LG, Sagittarius is the best-known example of a currently disrupting dwarf galaxy, but based on photometric studies tidal interactions have also been claimed for other dSphs such as Ursa Minor, Sculptor and Carina (e.g., \citet{1988AJ.....96.1352E, 1988AJ.....95.1706E, 1995MNRAS.277.1354I, 1996ApJ...469L..93K, 1997ASPC..127..163S, 2005AJ....130.2677M, 2000AJ....119..760M, 2001ApJ...549L..63M, 2003AJ....125.1352P, 2005ApJ...631L.137M, 2003A&A...406..847W, 2006AJ....131..375W, 2004ApJ...611L..21W, 2007ApJ...663..960S, 2006ApJ...649..201M, 2003ApJ...599.1082M,2008Natur.454.1096S}). Draco on the other hand, appears to be an example of an undisturbed dSph despite its proximity to the Milky Way (\citet{2001AJ....122.2538O,2003ApJ...589..798K}). Tidal interactions and their effects have also been extensively explored in N-body simulations (e.g., \citet{1999ApJ...512L.109J, 2002MNRAS.336..119M, 2003A&A...405..931P}). In this context, a general bimodal density profile can be used to describe the systems with an inner stellar population and a shallowly decreasing density profile in the outer regions as a consequence of the tidal interaction. But while the external density profile is generally attributed to the tidal interactions, it is unclear whether the inner regions of dSphs are strongly tidally influenced, or whether a large $M/L$ ratio can dampen the tidal gravitational shocks (see \citet{1997ApJ...474L..99B, 1997NewA....2..139K, 1999A&A...347...77G, 2002ApJ...566..838K, 2003ApJ...586L.123G, 2002MNRAS.330..792K, 2006AJ....131.2114W, 2006ApJ...642L..41W, 2007ApJ...663..948G})\footnote{A slightly different approach can be found in \citet{1995AJ....109.1071P} and \citet{1995ApJ...442..142O} where the authors span, in a specialized context, a wide range of eccentricities for exceedingly low-mass dwarf galaxies without dark matter (but see also \citet{2002ApJ...566..838K} and \citet{2003ApJ...589..798K}).}.
The radial velocity dispersion profile is a powerful tool for the investigation of multi-component self-gravitating systems, since it is sensitive to the dark matter distribution and accessible to observations (e.g. Pasetto et al. (2010)). Recently radial velocity data became available to track the line-of-sight velocity dispersion as a function of radius for many LG dwarf galaxies (e.g. \citet{2004ApJ...617L.119T, 2006AJ....131..375W, 2004ApJ...611L..21W, 2005ApJ...631L.137M, 2006AJ....131.2114W, 2006ApJ...642L..41W, 2007ApJ...663..960S, 2007ApJ...657..241K, 2007AJ....134..566K, 2006ApJ...649..201M, 2009AJ....137.3100W}) and these data permit a more detailed modeling of the dwarf galaxies kinematic status (e.g. \citet{1997NewA....2..139K, 1999AJ....117.1275K, 2004ApJ...608..663K,  2006MNRAS.367..387R, 2008ApJ...673..226P, 2009ApJ...698..222P}, Pasetto et al 2009). For the same approach to  multi-component dSphs in the cosmological $\Lambda$CMD
 context see, e.g., \citet{2002MNRAS.335L..84S} and \citet{2003ApJ...584..541H} or, in a purely dynamical context see e.g. \citet{2009MNRAS.393..179C}, Pasetto et al. (2010), where the primordial stellar population is embedded inside an extended dark matter halo.
 
In our current study, we concentrate on the Carina dSph galaxy, for which an extensive body of observational data is available from the literature. We explore the orbital evolution of this dwarf. 
The paper is divided in two main sections: Sect. \ref{Orbits01} reviews the observations on Carina and presents a preliminary orbit investigation in a point-mass approximation, while Sect. \ref{Nbodys} presents the N-body simulations and comparisons with the observations. Conclusions are presented in Sect. \ref{Conclusions}. 

\section{The family of orbits for the Carina dwarf galaxy} \label{Orbits01}
\subsection{Initial condition vs final condition}
In studying the orbits of the Carina dwarf galaxy, aspects of its actual dynamical and kinematic condition, as well as the history of its stellar populations and chemical enrichment, must be taken into account. The incompleteness of available data, together with their uncertainty or errors, combined with the limits of our analysis tools, increase the difficulties we have to overcome in order to produce a realistic model for Carina's orbit. Here, the complete amount of available data for this galaxy is too large to be completely reviewed, thus only the more salient characteristics are reported for the sake of brevity.

\subsection{Observations}
\subsubsection{Proper motions, distance and radial velocity}
The present observational data on the Carina dSph make it necessary for us to discuss families of orbits instead of a true single orbit. Here, we devote our main efforts to reducing the number of plausible orbits using all the available constraints known to date. We start with a determination of the proper motion in Table 1 of \citet{2008ApJ...680..287M} based on advanced charge transfer inefficiency (CTI) correction for the Space Telescope Imaging Spectrograph (STIS): $\left( {\mu _\alpha  \cos \delta ,\mu _\delta  } \right) = \left( { + 22 \pm 13, + 24 \pm 11} \right) \rm{mas \cdot yr^{ - 1}} $. We proceed with the reduction to an inertial galactocentric reference system $S_0 $, updating \citet{1987AJ.....93..864J} to J2000 as already done by e.g., \citet{2003A&A...405..931P}. We extensively describe the orientation  of the velocity space directions for this inertial reference system centered on the MW, $S_0$, in Appendix A. The resulting velocity vector for Carina is ${\bf{v}}_{{\rm{Car}}} = \left\{ { + 113 \pm 52, - 14 \pm 25, + 44 \pm 56} \right\}\rm{km \cdot s^{ - 1}}$. Proper motions for Carina have also been recently derived by \citet{2008ApJ...688L..75W} with error bars compatible with the values we adopted here.

The first determination for the distance of Carina comes from the RRLyrae-based distance modulus estimation by \citet{1986AJ.....92..302S}, and more recent values were determined by \cite{1994AJ....108..507S, 1998AJ....115.1840H, 1997AJ....114.1458M} and \citet{ 1995AJ....109.1751M}.

Measurements of the radial velocity were already made in an older work on carbon stars by  \citet{1983BAAS...15..907C}, where the authors deduced a radial velocity dispersion of $6 \rm{km \cdot s^{ - 1}} $. The actually value we assumed for the radial velocity is ${\rm{v}}_r  = 224\left( { \pm 3} \right) \rm{km \cdot s^{ - 1}}$ \citep{1998AJ....115.1856M, 1990ApJ...362L..55M}.

\textbf{\textsl{Methodology for the dwarf galaxy simulation:}}
We simulate a dwarf galaxy composed of three components: a dark matter component that dominates the total mass, a gas component and a stellar component. We assume that the preexisting old stellar population has settled after an initial violent-relaxation phase of the dark halo which created the potential well within which the baryonic component collapsed. The corresponding burst of star formation created the oldest stellar population that is observable through the analysis of the observed colour magnitude diagrams (CMD) (see next section). We assume a time-scale of a few Gyr for the realization of this initial 3-component dwarf galaxy model and its subsequent settling down in an orbit around the Milky Way (MW) with an initial orbital energy $E_{\rm{orb}}^{\rm{ini}} $. We start our simulations after this initial evolution has already taken place, i.e., we consider a dwarf galaxy in which the old population has already formed and that is already in orbit around the MW. We now consider the subsequent orbital evolution covering the last $ \cong 9$ Gyr. Hence we assume that the observed proper motions represent are indirectly related this initial orbital energy $E_{\rm{orb}}^{\rm{ini}} $. There is no reason to investigate the orbit prior to this initial time, because the system has still not settled down in thermodynamic/virial equilibrium and the concept of a barycenter for a complex of blobs of gas and stars is still meaningless.  Similarly, any old preexisting stellar population, by experiencing a collisionless violent relaxation process, loses memory of its initial phase-space distribution that, as a consequence, is completely unrelated to the currently observable kinematic properties of Carina (see following sections for a review of other possible scenarios of formation for Carina).

\subsubsection{Milky Way model}
In our simulation we included an external galaxy model resembling that of the Milky Way.  Nevertheless, considering the large Galactocentric distance of the dwarf galaxy we are going to analyse and the large uncertainties in the MW models, we will model the MW with a few simplified assumptions: 

\begin{enumerate}
	\item \textbf{Milky Way potential}. The Milky Way will be simplified as a vector field across which we will orbit first a point mass model of the Carina dwarf galaxy (in Sect. \ref{PMint}). Then, once we have reduced the plausible orbits suitable to our aims, we will orbit a full self-consistent N-body model of the Carina dwarf galaxy considering gravity, star formation and feedback processes of this 'live'-satellite.
We tested different potentials representing the MW in order to obtain results independent from our parametrization. Nevertheless, we want to stress that the large errors involved in the observational proper motion determination in general do not permit us to easily constrain the MW potential from the orbit determination of its satellites. 
The fully analytical parametrization is based on a logarithmic halo
\begin{equation}\label{an1}
\Phi _{\rm{h}} \left( {R,z,t} \right) = k_1^2 \log \left( {R^2  + k_2^2  + \frac{{z^2 }}{{k_3^2 }}} \right),
\end{equation}
for the disk we can choose a potential formulation as (after \cite{1975PASJ...27..533M})
\begin{equation}\label{an2}
\Phi _{\rm{d}} \left( {R,z,t} \right) =  - \frac{{GM_{\rm{d}} }}{{\sqrt {R^2  + \left( {k_4  + \sqrt {z^2  + k_5^2 } } \right)} }},
\end{equation}
and for the bulge
\begin{equation}\label{an3}
\Phi _b \left( {R,z,t} \right) =  - \frac{{GM_{\rm{b}} }}{{\sqrt {R^2  + z^2 }  + k_6 }},
\end{equation}
where the dependence on the position is made explicit by the cylindrical parametrization $(O,R,z)$ of $S_0$ while the dependence on the time $t$ is implicit in the parameter $k_i  = k_i \left( t \right)$ for $i = 1,...,6$ and on the total masses $M_{\rm{d}}  = M_{\rm{d}} \left( t \right)$, $M_{\rm{b}}  = M_{\rm{b}} \left( t \right)$ for the disk and the bulge respectively. $G$ is the gravitational constant $G = 6.673 \times 10^{ - 11} m^3 kg^{ - 1} s^{ - 2} $.
Particular attention is paid to the tuning of these parameters in order to match the nowadays known constrains on the MW potential. The adopted parameter are listed in Table \ref{TabellaAl1}. 

\begin{table*}
\caption{Parameters adopted as starting and final values for analytical MW potential of Eqn. \eqref{an1}, \eqref{an2} and \eqref{an3}.}
\centerline { \begin{tabular}{|lrrrrrrrr|}
\hline
$t_{\rm{ini}}$ &  $M_{\rm{b}}$         & $M_{\rm{d}}$         & $k_1$         & $k_2$ & $k_3$ & $k_4$ & $k_5$ & $k_6$     \\
			         &  $[M_\odot]$          & $[M_\odot]$          & [$km s^{-1}$] & [kpc] &       & [kpc] & [kpc] & [kpc]     \\
\hline
$Bulge$        & $ 3.4 \times 10^{10}$ &                      &               &       &       &       &       & 0.7       \\
$Disk$         &                       & $0.9 \times 10^{11}$ &               &       &       & 6.5   & 0.26  &           \\
$Halo$         &                       &                      & 120.8         & 10    &  1    &       &       &           \\
\hline
\hline
$t_{\rm{end}}$ &                       $                      $               &       &       &       &       &           \\
\hline
$Bulge$        & $ 3.4 \times 10^{10}$ &                      &               &       &       &       &       & 0.7       \\
$Disk$         &                       & $1.0 \times 10^{11}$ &               &       &       & 6.5   & 0.26  &           \\
$Halo$         &                       &                      & 130.8         & 12    &  0.8  &       &       &           \\
\hline
\end{tabular}}
\label{TabellaAl1}
\end{table*}

The evolution of the rotation curve for the radial range of Galacocentric distances of interest for the Carina's orbits is shown in Fig. \ref{RotCurve}. The upper blue thick line is e.g. compatible with the recent determination by \citet{2010arXiv1005.2619G} even if larger margin of error in the rotation curve is actually permitted in reltaion to the Sun location (e.g., \citet{2009ApJ...700..137R, 2009ApJ...704.1704B,2004AJ....127..914S}).

\begin{figure}
\resizebox{\hsize}{!}{\includegraphics{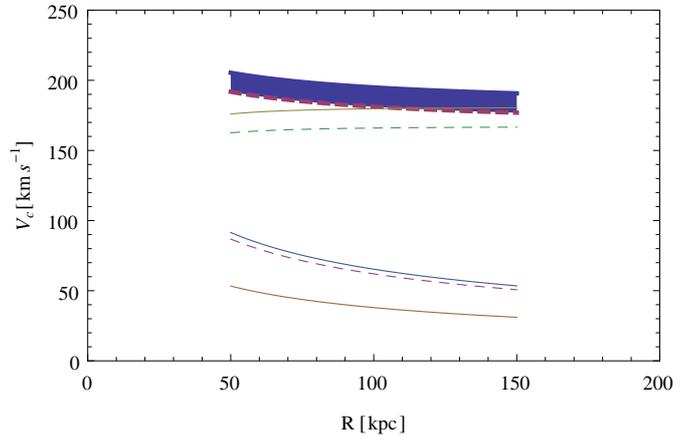}}
\caption{Rotational curve of the growing halo of the MW potential. The shadow represent the zone of gained mass for accretion in the outer part of the halo.}
\label{RotCurve}
\end{figure} 
We stress that different studies have found differences for the MW mass interior to the zone of relevance for our orbit integration. E.g., \citet{1996ApJ...457..228K} obtained $M_{tot} \left( {r < 100 \rm{kpc}} \right) = 7.5 \times 10^{11} \rm{M_ \odot}  $, and \citet{1998MNRAS.294..429D} obtained $M_{tot} \left( {r < 100 \rm{kpc}} \right) = 7 \times 10^{11} \rm{M_ \odot}  $. In addition, many different approaches can produce mass estimates for the MW, e.g. \citet{1998ApJ...500L.149Z} used the velocity distribution of the MW satellites, \citet{2003A&A...409..523R} used the escape velocity from the Local Standard of Rest, and \citet{1999Ap&SS.265..417C} from timing arguments. Similar results were found by \citet{2003A&A...397..899S} from the kinematics of Galactic satellites, halo globular clusters and horizontal branch stars. \citet{1995ApJ...439..652L} obtained $M_{tot} \left( {r < 100 \rm{kpc}} \right) = 5 \times 10^{11} \rm{M_ \odot}  $
 from the dynamics of the Magellanic Clouds, \citet{2004ASPC..317..203V} found $M_{tot} \left( {r < 100 \rm{kpc}} \right) = 8.8 \times 10^{11} \rm{M_ \odot}  $  from proper motion modelling of stars in two deep fields toward the North Galactic Pole. Mass estimate based on the moments of the Jeans equation are often sensitive to the anisotropy profile of the MW (e.g. \citet{2010AJ....139...59B}) leading to unsecured results, e.g. compare \cite{2005MNRAS.364..433B} and \citet{2006MNRAS.369.1688D}, or \citet{2010MNRAS.406..264W} that which quoted a value for $M_{tot} \left( {r < 300 \rm{kpc}} \right) \approx 1.4 \times 10^{12} \rm{M_ \odot}  $ and recently \cite{2008ApJ...684.1143X} found $M_{tot} \left( {r < 60 \rm{kpc}} \right) \approx 4.0 \times 10^{11} \rm{M_ \odot}$.

	\item \textbf{Milky Way accretion rate}. In these last 9 Gyr of evolution, we expect that not only the dwarf galaxy changed its physical properties, but that also the mass of the MW grew due to the continuous inflow of matter.
For simplicity we assume that 
\begin{itemize}
	\item The MW has not experienced any major merger in the last 9 Gyr,
	\item star formation processes are of minimal importance in the growth of MW.
\end{itemize}
These simplifications are realistic for our time range of interest of roughly $ \cong 9$ Gyr
 (e.g., \citet{2008MNRAS.384....2G}). What is most interesting for us is the evolution of the density distribution of the outer part of the halo, which is typically expected to be flatter than the inner part (see, e.g, \citet{2008A&ARv..15..145H} for a general review on this dichotomy), because the outer halo is where the orbit of a dwarf galaxy is expected to pass through and evolve. We represent the external mass distribution of the MW with an external potential that is time dependent. 

All these determinations lead to a wide range of values within which we want to choose our parameters when a growth factor is added in a time-dependent potential for the halo component. For the MW we assume that in the time range of interest, roughly the last 9 Gyr, the growth rate is mainly due to the minor mergers that are expected to be dominant in general for redshift $z < 1$ and for galaxies as massive as MW (\citet{2008MNRAS.384....2G}). Taking into account the uncertainties in the final target amount of dark matter, and the mass to light ratio of the MW as in \citet{2006MNRAS.372.1149F}, a simple linear growth factor of $10 \div 25\% $  is tested for the MW in the last 9 Gyr of evolution ($z < 1.46$ in redshift) (e.g. \citet{1999Ap&SS.265..417C, 2008MNRAS.384....2G}). Higher values for this parameter can be found in literature, although the relevence of the MW mass growing factor is limited by the large uncertentaintly in the total mass of the MW ($M_{tot}(r<100kpc)$). This uncertainties has unavoidable bad influence on our ability to determine the orbits of all the dwarf galaxies in the MW outskirts.
 we also test and take into account a gradual flattening of the inner halo, although there exist no data on Carina dwarf galaxy that are sensitive to this parameter: e.g. it can be proved, as already done in \cite{2003A&A...405..931P} that the presence of the bulge is not relevant for the integration of the orbit of Carina (similar as in the case of Sculptor in \cite{2003A&A...405..931P} and the following sections). The only observational data, e.g. from the SDSS \citet{2004AJ....127..899S, 2008ApJ...673..864J}, refers to the first 25 kpc of Galactocentric distance, that are extremly unlikely in the orbit of Carina (see Fig. \ref{FigOrbits01} and \ref{FigOrbits03} in the following section). 
	\item \textbf{Milky Way triaxiality}. The Milky Way potential and the possibility if its triaxiality are  investigated in different papers, e.g., \citet{2004ApJ...610L..97H}, \citet{2007ApJ...660.1264M}. Unfortunately the approach used in these studies is based on the possibility to track the orbits with the help of tidal streams, e.g. \citet{2008ApJ...683..750C, 2009MNRAS.400..548E, 2010ApJ...712..260K} and thus is limited to a few objects, mainly globular clusters or to the Sagittarius dwarf galaxy. Our target, the Carina dwarf galaxy,  does not show long tidal streams that would permit us to constrain the MW potential triaxiality.
	Hence, despite the simplicity of implementing a tiaxialhalo MW, we will not introduce here such a potential since its influence cannot be directly verified within the errors of our present work. 
	Moreover, triaxial halo presents challenges to the stability of the disks as shown in \citet{2010arXiv1006.0537K}, the origin and evolution of the dark matter axis ratios $c/a$ and $c/b$ are still unclear, and the relative orientations of the three axes are still a matter of debate.
Nevertheless, we stress that the role of triaxiality is invoked as fundamental tool to investigate the orbits of the MW satellites close to the galactic centre (e.g. \citet{2009ApJ...703L..67L}).

\end{enumerate}

\subsubsection{Star formation history}
In the literature, many authors have investigated the star formation history (SFH) of the Carina dSph since its discovery by \citet{1977MNRAS.180P..81C} and the pioneering work of \citet{1983ApJ...273..530M}. These studies demostrated the presence of multiple populations with different ages (\citet{1985A&A...144..388A,1986A&A...161..232A, 1994AJ....108..507S, 1990ApJ...362L..55M, 1997AJ....114.1458M, 1998AJ....115.1840H, 1990A&AS...82....1M}). \citet{2003AJ....126..218M}) published a deep CMD for the central zones of Carina, and \citet{2003ApJ...589L..85R} derived the SFH from a more spatially extended sample based on the analysis of RGB stars. We can then assume that the star formation in Carina consisted of an initial burst of star formation that formed the oldest stellar population, a second important burst around $ \sim 6$ Gyr ago, and a more recent event around $ \sim 2$ Gyr ago.
These episodes of star formation are clearly visible in Carina's CMD, and the existence of a fourth, even more recent episode has been proposed. What makes Carina special is that these episodes are recognizable as distinct events with distinct main sequence turn-offs.

\subsubsection{Chemical evolution}
Constraints on the Carina metallicity distribution came only recently from the spectroscopic studies or chemical evolution models (e.g. \citet{2006AJ....131..895K, 2006A&A...453...67L, 2007AN....328..652K, 2008A&A...481..161A, 2008AJ....135.1580K}). They generally show a quite spread in the metallicity distribution. We will consider the chemical evolution in more details in a separate work (Pasetto et al. in preparation).

\subsection{Point mass integrations}\label{PMint}
\begin{figure}
\resizebox{\hsize}{!}{\includegraphics{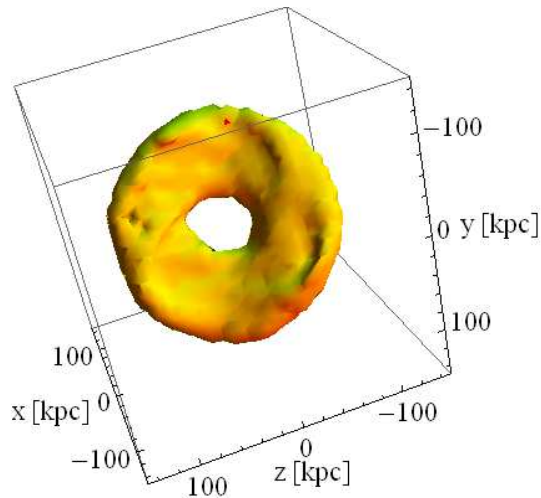}}
\caption{Error propagation for the point-mass approximation in the orbit of the Carina dwarf galaxy. We see the convolution surface for $10^5$ orbits. Although the uncertainty of the orbits does not permit us to go back in time for more than a few Gyrs, it is evident that a restricted set of values can be assigned to the orbital energy from the path of the galaxy in its first Gyr of backward evolution if we require to lie inside this manifold. This permits us to provide initial values for the full phase space description of the orbits that we want to integrate in a full N-body simulation, starting from the apocenter or the pericenter position. The use of these two different starting points as initial position of the barycenter of the gravitational system, although it refers to the same orbital energy, leads to different star formation histories.}
\label{FigOrbits01}
\end{figure}

\begin{figure}
\resizebox{\hsize}{!}{\includegraphics{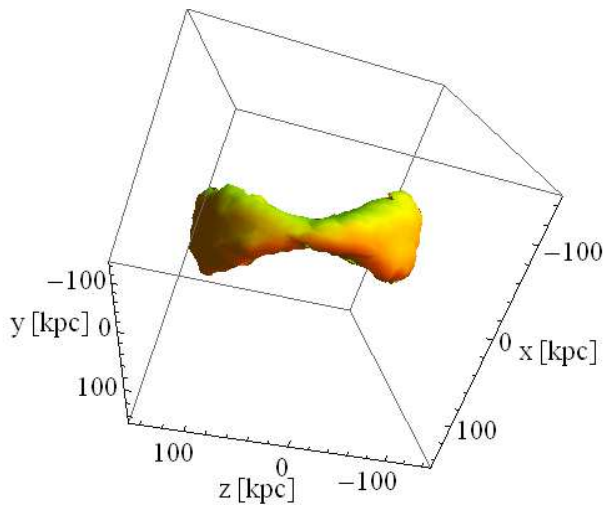}}
\caption{Same as Fig. \ref{FigOrbits01} but with an edge-on view. The box-type shape of the orbit and the permitted regions are evidenced.}
\label{FigOrbits02}
\end{figure}

\begin{figure}
\resizebox{\hsize}{!}{\includegraphics{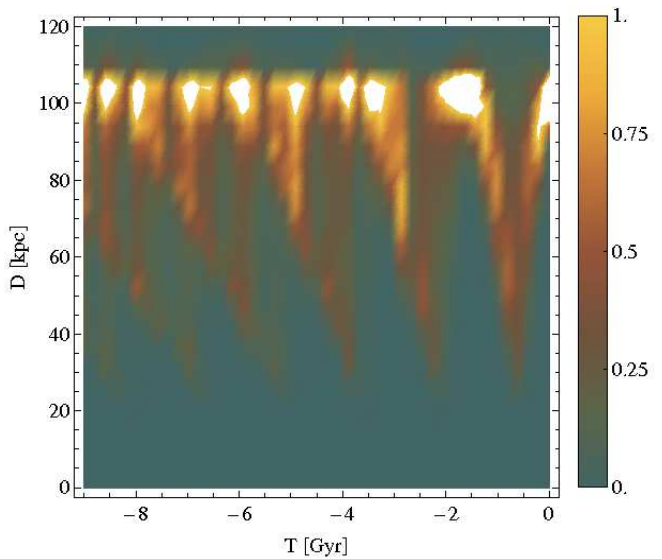}}
\caption{For the previous orbit integration shown in Figures \ref{FigOrbits01} and \ref{FigOrbits03}, we computed the radial distance of the dwarf galaxy barycentre in the respect to the inertial reference system centered on the MW. Here the probability of the location of the dwarf over the look-back time has been computed. Larger white zones (i.e., high probability) are closer to the present time where the indetermination has a minor impact. The indetermination affects mainly the pericenter passages where we have few or no constraints to satisfy due to the dominating effect of the MW growing halo potential.}
\label{FigOrbits03}
\end{figure}

Since our work is based on two full N-body integrators (e.g. \citet{1999A&A...348..371B, 1998MNRAS.297.1021C}) which can track simultaneously many different physical processes and produce results that are compatible with observational constraints, one might reasonably question the utility and validity of point-mass approximations. We contend that there are multiple advantages in the use of the point-mass approximation as a \textit{preliminary} study for orbit determination:
\begin{enumerate}
	\item The principal effect that can influence the orbit determination is the mass loss that the dwarf suffers during its pericentric passages. This mass change cannot easily be parametrized because it depends on the intensity of tidal effects as well as on the initial internal structure of the dwarf. Thus we will use phase-space coordinates and velocities obtained from point-mass calculations to get the initial values of the Carina orbit for the full N-body simulations only for the first 1 Gyr of backward evolution, i.e. before the time of the first pericentric passage $t_{p1}$. These initial guess values for the phase-space location of the barycenter of the N-body system span a full range of positions in configuration space from apocenter (where we will make an approximate match with the observations) to first pericenter (at the time $t = t_{p1} $) and lead to different star formation histories even if they correspond perfectly to the same initial orbital energy, $E_{\rm{orb}} \left( {\hat t} \right) = E_{\rm{orb}}^{\rm{ini}}  \equiv E_{\rm{orb}} \left( {t_0 } \right)\forall \hat t \in \left[ {t_0 ,t_{p1} } \right]$. As an example of the orbit indetermination resulting from the large error bars on the proper motion, we plot in Fig. \ref{FigOrbits01}, \ref{FigOrbits02}  and \ref{FigOrbits03} the orbit integrations, backward in the time $t$, for point-mass particles where initial values have been randomly sampled with uniform distribution inside the error bars for the velocities.
	\item The symmetries of the orbits remain evident. The expected appearance of the orbit of Carina from the point-mass integration is roughly a rosette-like orbit with several pericenter passages. This observation permits us to make some simplifying assumptions. The old stellar disk population of the MW is at least as old as 10 Gyr (e.g. \citet{2002ARA&A..40..487F}), thus we can assume that the gravitational potential of the disk and the bulge of MW has not changed its gravitational influence on the orbit of Carina. This allows to us to \textit{limit the growth rate in the MW mass only to the halo component, keeping fixed the amount of mass in the disk and the bulge}. A growing disk has nevertheless tested as well in our simulation as explained in Table \ref{TabellaAl1}.
  \item The rosette orbit symmetry reflects a time symmetry that permits us to backward-integrate the N-body gravitational system by simply changing the sign of the velocities. Nevertheless, the present structure to which we apply the orbital phase space and the external gravitational influence of the MW, describes a possible initial dynamical structure of the system as it was after the oldest stellar population had formed. It is often not considered in dynamical studies, in these kind of works, that the stellar evolution history does not have this time symmetry that is assumed in the pure dynamical approach. Clearly, the gas consumption and the chemical evolution are unidirectional also for the completely collisionless treatment of the dynamics. Thus the point-mass integration provides us with an initial volume of the phase space values to assign to a full N-body simulation model at the starting time, say $t = t_0  = 9$ Gyr ago.
\end{enumerate}

\begin{figure}
\resizebox{\hsize}{!}{\includegraphics{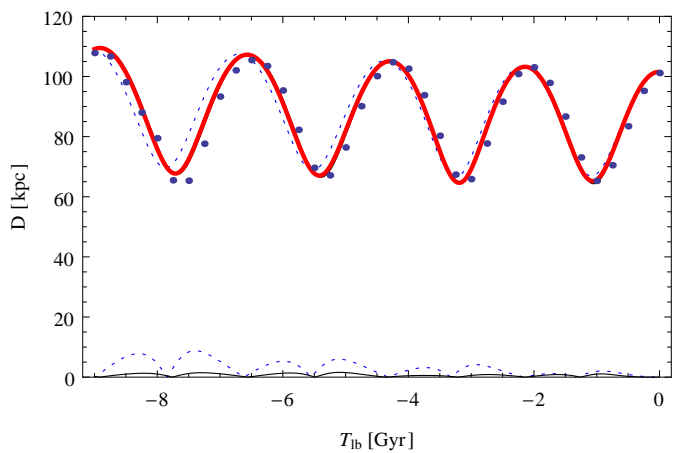}}
\caption{Radial distance evolution test. Here we compare the N-body simulation (black dots) and the point mass integration (red solid line). The dashed line represent the integration without the bulge commented in Sect. \ref{Robustnessofthepointmasssolutions}. The lines at the bottom represent the difference between the red-line and the dashed line, while the solid black line represent the difference between the red line and the orbit computed considering the dinamical friction (see text for details and Sect. \ref{Robustnessofthepointmasssolutions}).}
\label{FigNbodPointMass}
\end{figure}

Even though it is clear that the orbital plane determination of full N-body simulation and a point-mass integration in a spherical potential is supposed to be not too different between, we would like to show that this is also the case for the pericenter passages in this type of orbit.  This is done in Fig. \ref{FigNbodPointMass}. Here, we plotted the results of the point-mass integration for an orbit with a close pericenter passage for the set allowed by Carina's proper motion error bars. The same initial conditions were then evolved in a fully consistent N-body simulation. The results were obtained with two different codes: the first by \citet{1999A&A...348..371B} in a serial machine with special hardware dedicated (see \citet{SpurzemBMKLBMKB09}) and the second by \citet{1998MNRAS.297.1021C} on a parallel machine at the Juelich Super-computer Center. Both codes yeld comparable results across the entire range of initial values.

So why does our approximation work so well? 
If we consider the two galaxies MW and Carina, we have to treat them as extended bodies. In this case the acceleration of the center of mass of Carina is given by
\[
m_{\rm{I}}^{\rm{Car}} \frac{{d^2 {\bf{x}}}}{{dt^2 }} =  - \int_{}^{} {\rho _{\rm{G}}^{\rm{Car}} \left( {{\bf{x'}}} \right)\nabla _{{\bf{x'}}} \Phi _{MW} \left( {{\bf{x'}}} \right)d^3 {\bf{x}}} 
\]
where the integral is extended to the entire galaxy and we can no longer assume $\nabla _{{\bf{x'}}} \Phi _{MW} \left( {{\bf{x'}}} \right) \cong \rm{const}$ to take it outside the integral sign. This means that extended bodies orbit generally with the same velocities as a point particle if and only if  $\nabla _{{\bf{x'}}} \Phi \left( {{\bf{x'}}} \right) \cong \rm{const}$, i.e. if $\nabla _{{\bf{x'}}} \Phi \left( {{\bf{x'}}} \right)$ is roughly constant within the body volume of the Carina dwarf galaxy and it can be extracted off the integral sign to obtain the point mass equation of motion. In general, this approximation cannot be applied to galaxies because they are neither bodies that could be considered small enough such e.g., in the case of a globular cluster orbiting around the central zones of the bulge, nor can they move in a sufficiently homogeneous gravitational field e.g., they can reach the inner zones of the MW galaxy where the bulge influence become dominant (see  Sect. \ref{Robustnessofthepointmasssolutions}).  Using the consideration on the MW potential given in the previous section we have taken care to check that the orbital range of Carina in the external potential of the MW satisfies the required approximation in every orbit computation performed with the point-mass approximation and in minimizing the Action (see next section). Thus we safely proceed further with the approximation $m_{Car} \left( t \right) \cong m_{Car}  = const$ and the point-mass approximation in the next section where we develop a dedicated method to select a preferential initial condition for the full N-body simulations. In the N-body simulations explained in the next section mass loss effects will be self-consistently taken in consideration.

\subsection{Minimizing the action}\label{Minact}
After these preliminary considerations it became evident that, although the the errors in the proper motions are quite large, at least the orbital plane can be inferred relatively easily from the orbit analysis. 

For any chosen value within the range permitted by the error bars of the proper motions, the orbital plane remains roughly determined. The range of orbits allowed by the error bar permits us to deduce that the orbits are more likely box-type. \textit{That results not from the orbit determination which is integrated for too short a time range, but from the consideration of full sample of orbits integrated}. 

The true orbit, is probably contained in this plane but we want to proceed by constraining the possible orbits further with an alternative approach. The approach of using a point mass for the orbiting dwarf galaxy can bias the determination of the pericentre and apocentre of the orbit evolution if the dwarf passes very close to the inner zones of the galaxy. Nevertheless, we do not expect it to influence the plane determination of the orbit, since the Carina dwarf galaxy is orbiting mostly in the external regions of a growing MW halo, nor do we expect to it to affect the error estimates of the pericentre position to be so large as to invalidate the following considerations. To support these approximations we remember that when Carina is more massive (before the first pericenter passage) the MW halo still has its spherical shape, which is gradually converted into an ellipsoidal shape at $t_{end}$ with $k_3$ parameter changing in Table \ref{TabellaAl1}) from 1 to 0.8. Based on Fig. \ref{FigOrbits03} we can prove that the probability that Carina intersects the disk of the MW is null and the probability of a pericenter lower than 60 kpc is only a few percent!

The generic Lagrangian of the dynamical MW-Carina system can be written down as 
\begin{equation}\label{Lagrangian}
	L = \frac{1}{2}m_{\rm{Car}} \left\| {{\bf{\dot x}}\left( t \right)} \right\|^2  - m_{\rm{Car}} \Phi _{\rm{Gal}} \left( {{\bf{x}},t} \right).
\end{equation}
The Lagrangian is a function that lives in the tangent bundle of the configuration manifold $\mathbb{Q}$ of Fig. \ref{FigOrbits01} or \ref{FigOrbits02}, thus it is defined in $L:T\mathbb{Q} \times \mathbb{R} \to \mathbb{R}$ (where $\mathbb{R}$ is the infinite time line parametrized by the time $t$) and its integral can be easily solved numerically. Thus, we ask ourselves: what is the orbital plane within the error bars of the proper motion that minimizes the action? The principle of minimal action will lead us to possible stationary points for the integral of the Lagrangian. Hence, we proceed by minimizing the action integral $S\equiv\delta \int\limits_{t_0 }^0 {Ldt} $, subject to the constraints on the initial value that we deduce from the proper motion and the star formation history. 
We can, in fact, \textit{suppose }that the episodes of star formation have been triggered by the tidal influence of the MW potential in the pericentre passages (the \textit{proof} of this hypothesis will come only in the next section). Although in $T\mathbb{Q}$ the equations of motion are a set of \textit{first}-order differential equations (i.e., through each point $\left( {{\bm{x}},{\bm{\dot x}}} \right)$ of $T\mathbb{Q}$ passes just one solution of the equations of motion) thus permitting us to easily integrate the action $S$, before performing a minimization of the action 
we can already restrict the space $T_{{\bm{x}}_{{\rm{Car}}}(0) } \mathbb{Q}$, i.e. the fibers (in the geometrical sense) of the Carina configuration position at the initial time ${\bm{x}}_{{\rm{Car}}}(0)$. 

\begin{figure}
\resizebox{\hsize}{!}{\includegraphics{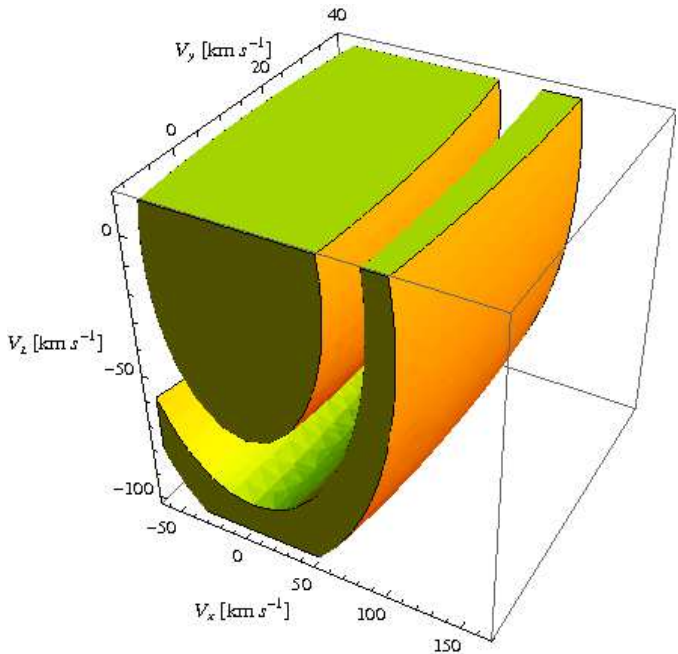}}
\caption{$T_{{\mathbf{x}}_{{\text{Car}}} } \mathbb{Q}$ at $t=0$. The section provides the zones within which the condition expressed by Eqn. \eqref{conditionMin} is satisfied. This condition clearly represents a good constraint when compared with the whole space where we want to find the minimum of the action. The advantage is in the more restricted number of points that we have to sample, for which we have to integrate the differential equation and to solve the action integral as well as in the reduced computational time. The volume of the velocity space to sample in this figure is more than 1/5 smaller than the overall space without the constraints imposed from the star formation history.}
\label{VelSpace}
\end{figure}

In terms of Newtonian dynamics, we already gain a more restrictive velocity space than $
V^3  \equiv \left\{ {\left. {\bm{v}} \right|{\bm{v}} \in \mathbb{R}^3  \wedge {\bm{v}} - {\bm{\sigma }}_{\bm{v}}  \leqslant {\bm{v}} \leqslant {\bm{v}} + {\bm{\sigma }}_{\bm{v}} } \right\}$
 if we exclude from our velocity space all the orbits that do not lead to a pericentre passage at the times $t_1, t_2 $ corresponding to the two bursts of star formation at $t_2  \sim 6$ Gyr and $t_1  \sim 2$ Gyr. Moreover, we do not pretend to know the precise moment for the time of closest approach, but we can provide the time range $\delta t_i $, $i = 1, 2$ (e.g. \citet{1998AJ....115.1840H}). In general we expect $\delta t_2  > \delta t_1 $  and both not to be smaller that 1 Gyr. The desired conditions we want to implement in our minimization tool can easily be achieved by requiring a minimum in the distance function $\left. {d\left( {t,{\bm{v}}_0 } \right)} \right|_{\rm{Car}} :\mathbb{R} \times V^3  \to \mathbb{R}$ in the range $\left[ {t_1  - \frac{{\delta t_1 }}
{2},t_1  + \frac{{\delta t_1 }}
{2}} \right]$ and in the range $\left[ {t_2  - \frac{{\delta t_2 }}
{2},t_2  + \frac{{\delta t_2 }}
{2}} \right]$: 
\begin{equation}\label{conditionMin}
	\left. {\partial _t d\left( {t,{\bm{v}}_0 } \right)} \right|_{t = \hat t}  = 0 
 \wedge  \left. {\partial _{t,t} d\left( {t,{\bm{v}}_0 } \right)} \right|_{t = \hat t}  > 0
\end{equation}
Here clearly the continuity of the partial derivative with respect to the time $\partial _t $ is assumed. We underline immediately here that, as evidenced by Eqn. \eqref{conditionMin}, this condition \textit{does not require at all to exclude other pericentre passages}. 
In this case of the manifold at $T_{{\mathbf{x}}_{{\text{Car}}} } \mathbb{Q}$ at $t=0$ results in the plot shown in Fig. \ref{VelSpace} as an initial velocity space.

At this stage, here we \textit{assume} here that the tidal field acts as a trigger of the star formation. The proof of this assumption will be provide later with a selfconsistent generation of the star formation history of Carina in the context of the smooth hydrodynamic particles approach integrated in a full N-body simulation.

Nevertheless, by excluding those orbits that do not show any pericentre passage in the distance range required by the matching of Carina's SFH will reduce further the space where we have to minimize the action (and the computational time required for the minimization!). We adopt the same methods of minimization as \citet{2007A&A...463..427P}. We point out that the method applied here does not require any previous orbit parameterization as in a cosmological context (e.g., \citet{1989ApJ...344L..53P}). Moreover the alternative approach that comes from the Fourier series decomposition used in stellar spectral dynamics (e.g. \citet{1998MNRAS.298....1C}) cannot be exploited here because of the small number of orbital periods during which we can integrate the dwarf galaxy (no more than 9 Gyr and an orbital period between 2 and 4 Gyr) could lead to wrong frequency determinations. Nevertheless, for this small number of periods the full numerical integration of the action works fine. The only limitation of our method is the treatment of the mass loss in Eqn. \eqref{Lagrangian}: an approach for a dwarf galaxy experiencing a strong dwarf-bulge interaction as e.g., in the case of  Sagittarius, is not able to reproduce correctly the orbits we obtain with a full N-body simulation. This limits our methodology to the more distant dwarfs such as the Carina dwarf galaxy.

With this approach, we can generate a the family of orbits with the following values (obtained through the minimization process of the action for the Carina dwarf galaxy starting at $t_{lb}  = 9.645 \rm{Gyr}$):
\begin{equation}\label{PSini}
	 \begin{gathered}
  {\bm{x}}_{\rm{Car}}  = \left( { - 73, - 64,3} \right) \rm{kpc,} \hfill \\
  {\bm{v}}_{\rm{Car}}  = \left( {65, - 115,75} \right) \rm{km \cdot s^{ - 1}.}  \hfill \\ 
\end{gathered}  
\end{equation}
These are the initial phase-space values for most probable volume of $T_{{\bm{x}}_{{\rm{Car}}}(t_{lb}  = 9.645 \rm{Gyr}) } \mathbb{Q}$ for the galaxy at the initial time of our N-body simulation. 

After a full N-body simulation run, the previous initial conditions lead to the following present-day best fit values (values for the barycentre of the galaxy in a full N-body integration at $t_{lb}  = 0$):
\begin{equation}\label{PSfin}
	 \begin{gathered}
  {\bm{x}}_{\rm{Car}}  = \left( {22,89, - 34} \right) \rm{kpc} \hfill \\
  {\bm{v}}_{\rm{Car}}  = \left( { - 136,13, - 45} \right) \rm{km \cdot s^{ - 1},}  \hfill \\ 
\end{gathered}  
\end{equation}
(values in the reference system  $S_0$) which are fully compatible with the proper motion, radial velocity and distances that are actually observed.

\subsection{Robustness of the point mass solutions}\label{Robustnessofthepointmasssolutions}
The robustness of the the point mass integration is the subject of many cosmological or pure dynamical studies but often without a self-consistent consideration of the star formation processes (e.g., recently \citet{2005MNRAS.364..977P, 2010MNRAS.tmp..842L}). The relation between star formation and orbit determination is first time fully exploited here as key ingredient to constraints the Carina orbit determination. Nevertheless, once we have determined a family of solutions as in Eqn. \eqref{PSini}, and before generating a complete N-body integration that self-consistently considers the structural parameters of Carina and its star formation history, we check this family of orbits against the possible role that the dynamical friction could have played  \citep{1943ApJ....97..255C} and the parameters of the potential.  

\subsubsection{The dynamical friction}
In the last century, the perturbative methods applied to dynamical friction studies have achieved several important results in order to overcome the limitation of the original formulation in the celebrated article by Chandrasekhar (i.e. mainly the isotropy of the surrounding stellar field, see \citet{1984MNRAS.209..729T, 1986ApJ...300...93W}). Closely related to this approach is also the work of \citet{1992ApJ...390...79B,1999MNRAS.306....1N} that related the dynamical friction to the fluctuation-dissipation theorem. We will follow the treatment presented in the paper of \citet{1998ApJ...502..150C} which contains as special limit the Chandrasekhar work (e.g. \citet{1983Ap&SS..97..435K}). In this formulation of the perturbative theory (extension of previous works of e.g., \citet{1981ApJ...244..316K, 1994A&A...290..709S, 1996A&A...310..757S}) the authors recollect and extend several important results in order to overcome the limitation of the Chandrasekhar formula allowing one to follow analytically the gravitational wake influence, the tidal deformation and the shift of the barycentre of the primary galaxy. Moreover, this formulation unifies the local and large-scale interpretation of dynamical friction and provides a treatment even superior than the full N-body simulations often limited by the number of particles (see \citet{1999ApJ...525..720C}). What is more interesting for us, is that in this last paper the authors already proved that the role of the dynamical friction for the MW dwarf satellite is marginal, especially with reference to the more distant dwarfs, and in conflict with what would be expected from the previous straightforward application of the Chandrasekhar formula. 
Finally, if we want to perform a simple check, under the approximation presented in the paper of \citet{1999ApJ...525..720C} with the further simplification that comes from the constraint on previous investigated relevant pericenter distance at which the orbit of Carina is expected to evolve have its pericenter distance, we can write rewrite the equation of motion in a system, say $S_1$, as
\begin{equation}\label{eqsiF}
\mu \frac{{d^2 {\bf{x}}\left( t \right)}}{{dt^2 }} =  - GNm\frac{{{\bf{x}}\left( t \right)}}{{\left\| {{\bf{x}}\left( t \right)} \right\|^3 }}\int_{r' < {\bf{x}}\left( t \right)}^{} {d^3 {\bf{x}}\rho \left( {r'} \right) + {\bf{F}}_\Delta  }, 
\end{equation}
where $G=4.4926 \times 10^{ - 6} {\rm{kpc}}^3 {\rm{Gry}}^{ - 2} M_ \odot ^{ - 1} $ is the gravitational constant, $\mu  = \frac{{M_{{\rm{MW}}} M_{{\rm{Car}}} }}{{M_{{\rm{Car}}}  + M_{{\rm{MW}}} }}$ the reduced mass of Carina and MW, $\rho $ the halo density profile and ${{\bf{F}}_\Delta  }$ the back reaction force as presented in Appendix of \citet{1999ApJ...525..720C}. An alternative faster integration, which given the distance of Carina is in our case absolutely equivalent in its results, is based on a time dependent Coulumb logarithm $\ln \Lambda  \sim \ln \left\{ {\frac{{r_{apo}  - R\left( t \right)}}{\varepsilon }} \right\}$ with $\varepsilon  \sim \frac{{GM}}{{V^2 }}$ (see \citet{1976MNRAS.174..467W,1989MNRAS.239..549W}) used in the the infinite, homogeneous, non-self-gravitating collisionless background formulation by Chandrasekhar:
\[
{\bf{F}}_\Delta   =  - 4\pi \left( {GM} \right)^2 \rho \ln \Lambda \left( {{\mathop{\rm Erf}\nolimits} \left( x \right) - \frac{{2x}}{{\sqrt \pi  }}e^{ - x^2 } } \right)\frac{{\bf{V}}}{{\left\| {\bf{V}} \right\|}},
\]
where $x=\frac{{\left\| {\bf{V}} \right\|}}{{\sqrt 2 \sigma }}$ and ${\bf{V}}$ si the velocity vector of Carina.

The application of this formalism to our family of orbits for Carina confirms that the role of dynamical friction on our orbit determination is irrelevant as shown in Fig. \ref{FigNbodPointMass}. The orbit computed with Eqn. \eqref{eqsiF} almost precisely overlap the orbit determined without the dynamical friction (Sect. \ref{Minact}) thus we plotted a the bottom of the distance between the two orbits with a thin solid black line. We stress for comparison, \textit{that the error in the knowledge of the present location of the Carina dwarf galaxy is of about 10 kpc in radial distance} (e.g. \cite{1998AJ....115.1856M}).

\subsubsection{The MW potential}
The role of the MW potential is fundamental in evaluating the orbit of Carina. Recently \citet{2009ApJ...703L..67L} proved the necessity to fully exploit the triaxiality of the MW potential in order to explain the tidal streams location and the orbit of Sagittarius.
Unfortunately the situation for Carina is much more complicated. The error in its present location and the absence of long tidal streams that can somehow track the orbits (or in the contrary can constraint the MW potential) prompted us to apply a different technique taking the star formation history, as a new constraint on the orbit. I.e., we are explicitly \textit{assuming} that Carina's unusual star formation history was influenced by encounters with the MW.

As an illustrative example we point out here how for the Carina orbit integration the existence whole presence of the bulge ($\sim 10^{10} M_ \odot  $) can be neglected! Carina is orbiting in zones where the relative influence of the potential is mainly dominated by the halo and partially by the disk: ${{\Phi _{{\mathop{\rm bulge}\nolimits} } \left( {{\bf{x}},t} \right)} \mathord{\left/
 {\vphantom {{\Phi _{{\mathop{\rm bulge}\nolimits} } \left( {{\bf{x}},t} \right)} {\Phi _{{\mathop{\rm tot}\nolimits} } }}} \right.
 \kern-\nulldelimiterspace} {\Phi _{{\mathop{\rm tot}\nolimits} } }}\left( {{\bf{x}},t} \right) < 0.1$  for every point ${\bf{x}}$ and instant $t$ of interest in our orbit determination. By using the best fit orbit determination parameters (see Sect. \ref{Minact} Eqn. \eqref{PSini}) in Fig. \ref{FigNbodPointMass} we plotted with the dotted blue line the orbit distance computed without the bulge. As evidence in the bottom of the plot, the distance between this orbit and the red orbit for our best solution retains a value comparable with the error bar as large as the present day Carina distance uncertainty (10 kpc). This result shows that the role of the bulge is marginal in the orbit of Carina.

\section{N-body integration}\label{Nbodys}
The search for the best initial conditions of the orbital parameters of Carina must include also the matching of the internal properties of the dwarf. To accomplish this task we performed a large number of simulations in the space of the the restricted initial orbital condition that surround the best fit values obtained from the minimal action in the previous section. 

In order to perform this analysis in an automatic way, we wrote a code that computes all of the desired properties (derived star formation history, chemistry and kinematic properties of the orbiting evolved object) centred on the non-inertial barycentre of the particles representing the stellar component of the dwarf, say $S_2$. The barycentre of the moving dwarf's stellar component also has to be automatically computed to speed up the analysis of the dwarf's orbits. This is performed with an original method based on the down-hill simplex algorithm (explained in Appendix \ref{ApBar}).

A detailed description on the implementation adopted for the star formation recipes, chemical evolution and dynamics for this GRAPE-treeSPH code can be found in \citet{1999A&A...348..371B, 2003Ap&SS.284..865B, 2003CoKon.103..155B, 2002Ap&SS.281..297B,2001KFNT...17..213B,2000KFNTS...3...91B} and references therein. Special purpose hardware (\citet{SpurzemBMKLBMKB09}) was used in order to carry out a large number of high-resolution chemo-dynamical simulations.

\subsection{Initial conditions}
 The structural properties were adopted from an extension of the isolated models presented in \citet{2010A&A...514A..47P} in order to include a larger set of scale length initial conditions (Pasetto et al. 2010 in preparation). \textsl{Those isolated models then are evolved in the orbits around the MW selected as explained in the previous section. Star formation history, chemistry and dynamical properties of the Carina dwarf galaxy are self-consistently determined along the 9 Gyr of evolution.}
 
The staring initial condition can be partially tuned with the help of the literature results for Carina. Color magnitude diagrams were studied in \citep{ 1998AJ....115.1840H,1997AJ....114.1458M,1992PASAu..10...83M,1990A&AS...82..207M,1990A&AS...82....1M,2006MmSAI..77..270M,2006cams.book..272M,2004ASPC..310..133M,2004MSAIS...5...65M,2004MmSAI..75..114M,2003AJ....126..218M,2003MmSAI..74..909M,2004MmSAI..75..110R,2003ApJ...589L..85R}). Spectroscopic measurements yielded chemical abundances, e.g., \citet{2006AJ....131..895K,2008AJ....135.1580K}, the M/L ratio, e.g., \citet{2007ApJ...663..948G} and the dark matter distribution, e.g., \citet{2003ApJ...584..541H, 2007ApJ...663..948G}. From the analysis of the CMD of Carina (e.g. \citet{1998AJ....115.1840H}) we inferred that the oldest stellar population contributes no more than 15\% of the stellar mass and from the work of \citet{1993AJ....105..510M} we got a mass to light ratio in the V band of $\left( {\frac{M}{L}} \right)_V  \sim 23\left( {\frac{{M_ \odot  }}{{L_ \odot  }}} \right)_V $. Thus from Table 3 of \citet{2003ApJ...584..541H} we got $L = 0.4 \cdot 10^6 L_ \odot  $, which leads to a stellar mass estimate of $M \approx 23 \cdot 0.4 \cdot 10^6 M_ \odot   = 9.8 \cdot 10^6 M_ \odot  $. If we now assume that the mass lost by tidal interactions could reach an order of 95\%, following e.g., \citet{2003ApJ...584..541H} we get $M_{\text{Car}}  = 2 \cdot 10^8 M_ \odot$ and with an intial dark matter contents at least 50 times the baryonic fraction \citet{1998AJ....115.1840H}. The baryonic material consists of 15\% stars, 85\% gas. We thus adopted a starting value of $M_{\text{gas}}  = \frac{{M_{\text{dark}} }}{{50}}\frac{{85}}{{100}} = 3.3 \cdot 10^6 M_ \odot  $ and $M_{\text{star}}  = \frac{{M_{\text{dark}} }}{{50}}\frac{{15}}{{100}} = 5.9 \cdot 10^5 M_ \odot  $ for our reference model. These numbers can be considered as starting values for an orbiting dwarf galaxy, aroung which we want to find our best fit initial values. The final best matching initial condition are reassumed in Table \ref{Tabella01}. The models are represented by a 3-component self-gravitating system where the family of the multi-gamma models was  adopted for the density profiles (e.g. \citet{1993MNRAS.265..250D, 1994AJ....107..634T}). The scale parameters for the initial models were explored around the values of $r_c  \cong 0.72$ for the gaseous and stellar components and of $r_c  \cong 2.2 \rm{kpc}$ with a cut off around $20 \rm{kpc}$ for the dark matter component. The initial inner slope of the density profile was chosen around $\gamma  = 2$ to produce a massive and cuspy dark matter profile, while a more shallow profile with $\gamma  = 3/2$ was adopted for the gas and stellar profiles.

Considered the preliminary results on the orbit analysis from Sect. \ref{Minact}, models with an initial cuspy density distribution for the dark matter component have to be preferred in order Carina to survive four pericenter passages. Models with a starting relevant amount of gas suffer a flattening phenomenon of the DM density profile by the star formation processes as evidence in \cite{2010A&A...514A..47P}. Flatter DM density profiles have a reduced probability to survive different pericenter passages (e.g., \citet{2010MNRAS.406.1290P}).

  The initial temperature for the gas can be inferred assuming a spherical collapse model for the initial dark matter model and a matter power spectrum $P(k)$ at redshift zero compatible with studies from the SDSS, e.g., \citet{2004ApJ...606..702T} or work on the analysis of the Lyman-$\alpha$ forest, e.g., \citet{2002MNRAS.334..107G}. In this case the typical RMS internal velocity of a halo within $M \le 10^8 M_ \odot  $ is less than $13{\rm{km}} \cdot {\rm{s}}^{ - 1} $ and therefore we assumed an initial sound speed in hydrogen of roughly around $c_s  \cong 10{\rm{km}}\cdot{\rm{s}}^{ - 1} $ corresponding to a temperature of $T \cong 10000^\circ \text{K}$.

Starting with a three-component model, we implicitly assume our evolution to start after the oldest stellar population settled to form the central nucleus of the Carina dwarf galaxy. Any burst of star formation that occurred earlier than 9-10 Gyr ago cannot help us to trace back the orbital parameters. Moreover, the timescale permitting to the initial proto-cloud to collapse, in order to form the initial old population, and finally to settle on the orbit, requires no less than 3 Gyr (according to e.g., the chemical enrichment models \citet{2006MNRAS.371..643M}). Thus our orbit determination will start on a possible Carina dwarf galaxy configuration of 9 Gyr ago.

\subsection{Match with the observations}
The dynamical analysis we present shows the best-fitting model for the data as compared to the observation of the surface density profile of \citet{1995MNRAS.277.1354I}, and more recently updated in \citet{2006ApJ...649..201M} or \citet{2007ApJ...667L..53W} with a radial velocity dispersion profile.  The assumed initial and final best-fit model is presented in Table \ref{Tabella01}.

With the numerical scheme adopted the number of particles is \textit{not} constant (see, e.g. \citet{2010A&A...514A..47P}, \citet{1999A&A...348..371B},  \citet{2000A&AT...18..829B} and references therein for detailed formulation of the star formation processes). The starting models are described with 350000 particles growing with time up to half a million of particles, and the reference model we are using for the match with the present day observations (after 9 Gyr of evolution) contains about $1.2 \times 10^5$ stars within the 5 kpc from the barycentre of the Carina dwarf galaxy.

\begin{table}
\caption{Parameters for the galaxy model evolved in orbits around the potential. Effective radius and central velocity dispersion are omitted because the whole extent of the profile is matched (see the text for further details). The model is set at the starting time $t=t_0$ in the position of the phase space of Eqn. \eqref{PSini} and reaches the phase-space point given by Eqn. \eqref{PSfin}. The location in galactocentric coordinates is also in excellent agreement with the observation (e.g. \cite{1998AJ....115.1856M})}
\begin{tabular}{|l|rr|}
\hline
                    &  $t_{lb}=t_0$                    & $t_{lb}=0 $                   \\
\hline
l										& 																 & $\cong 260$ [deg]  \\
b										&																	 & $\cong -22$ [deg]  \\
d										& 																 & $\cong 100$ [kpc]  \\
$V_{l.o.s.}$        &																	 & $\cong 219$ [$km s^{-1}$] \\
$M_{\rm{gas}}$      & free parameter                   & $ \cong 0$                    \\
$M_{\rm{star}}$     & $1.998 \times 10^7 \rm{M_ \odot} $ & $\cong 1.9 \times 10^6 \rm{M_ \odot} $  \\
$M_{\rm{dark}}$     & $1.978 \times 10^8 \rm{M_ \odot} $ & $0.675 \times 10^8 \rm{M_ \odot}$   \\
\hline
\end{tabular}
\label{Tabella01}
\end{table}

\begin{figure*}
\resizebox{\hsize}{!}{\includegraphics{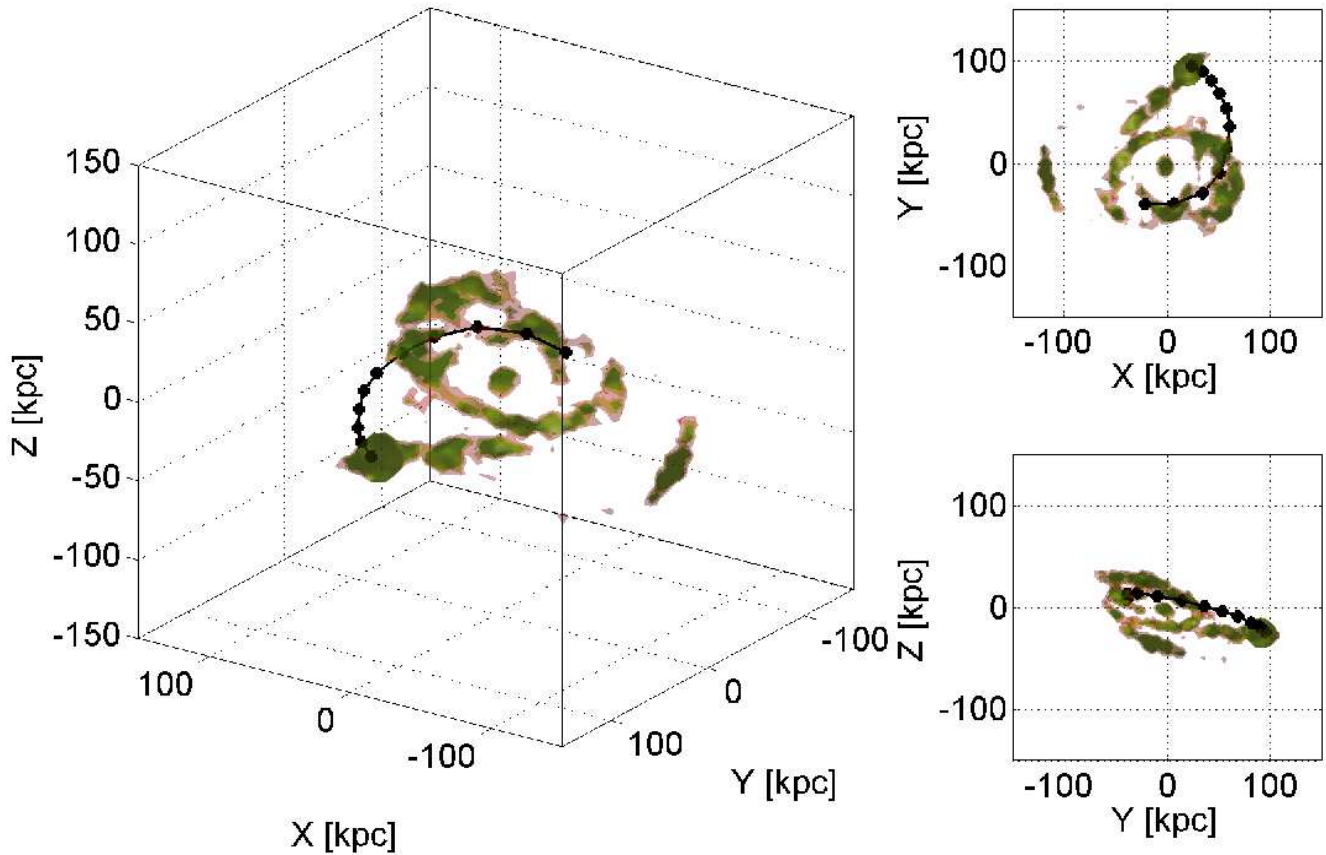}}
\caption{N-body representation of the evolved galaxy. The classical point-representation has been substituted with an 3D isocontour plot by counting the amount of stars within a 3D grid spanning all the configuration space of the orbit and joining cell with equal number of stars. The green isocontours show overdensities of $1/10^4$ and the purple for overdensities of $1/10^3$.}
\label{BellaFigura01}
\end{figure*}

\begin{figure}
\resizebox{\hsize}{!}{\includegraphics{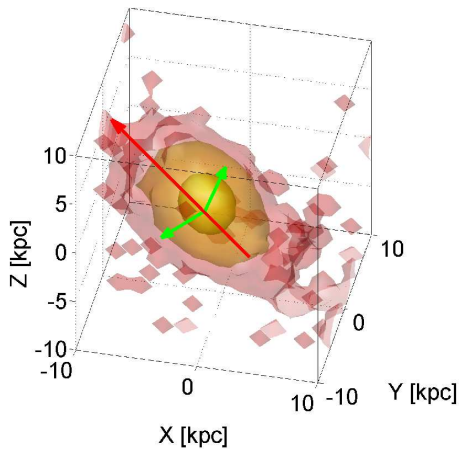}}
\caption{A zoom of the previous figure centred on the dwarf galaxy system $S_2$. The red arrow shows the eigenvector direction pointing in the radial direction where the radial profiles are computed (see text for details). \textit{We have not performed our analysis in comparison with the observation along the l.o.s. thanks to the spherical symmetry developed by of the system at t=0.}}
\label{NbodFig02}
\end{figure}

An illustrative sketch of the evolution of the system after about 9.6 Gyr of evolution is presented in Fig. \ref{BellaFigura01}. The plot shows several low resolution streams where the mass density drops down to ${\raise0.7ex\hbox{$1$} \!\mathord{\left/
 {\vphantom {1 {10^3 }}}\right.\kern-\nulldelimiterspace}
\!\lower0.7ex\hbox{${10^2 }$}} \div {\raise0.7ex\hbox{$1$} \!\mathord{\left/
 {\vphantom {1 {10^5 }}}\right.\kern-\nulldelimiterspace}
\!\lower0.7ex\hbox{${10^3 }$}}$ of the total mass thus showing clearly how the tidal tails developed by the Carina dwarf galaxy are expected to be extremely diffuse after $9$ Gyr of evolution (i.e. now). In this sense the more diffuse zones have characteristics that the model can predict but where the observations are lacking or their interpretation is insecure, e.g. \citet{2005AJ....130.2677M}.
The last snapshots of the orbit evolution are also presented with a black solid line. We observe the general 'tidal tail flipping' phenomenon already evidence in high resolution simulation by e.g., \citet{2009MNRAS.400.2162K}. The model at a generic snapshot is analyzed in a reference system centered on the baryonic matter barycenter as can be seen in Fig. \ref{NbodFig02}. 

We use Fig. \ref{NbodFig02}, to illustrate how the tidal tails developed in the simulation since the first pericentre passage, allowing us to define an ellipsoid as soon as we plot the 3D contours for even more diffuse regions. 
In Fig. \ref{NbodFig02} a yellow ellipsoid is visible in the center for which we can easily compute the inertia matrix to define its principal axis. Once the principal axes along the tidal tail direction are determined, we can perform our analysis along this radial direction (e.g., the red arrow in this figure). However, we underline that the radial \textit{stellar }profiles in the orthogonal principal directions present similar profiles not evidencing any pronounced ellipsoidal shape at $t=9.6 Gyr$ (thus providing a good agreement with the observational approach used in \citet{2006ApJ...649..201M}).

\subsubsection{Surface density profile}
\begin{figure}
\resizebox{\hsize}{!}{\includegraphics{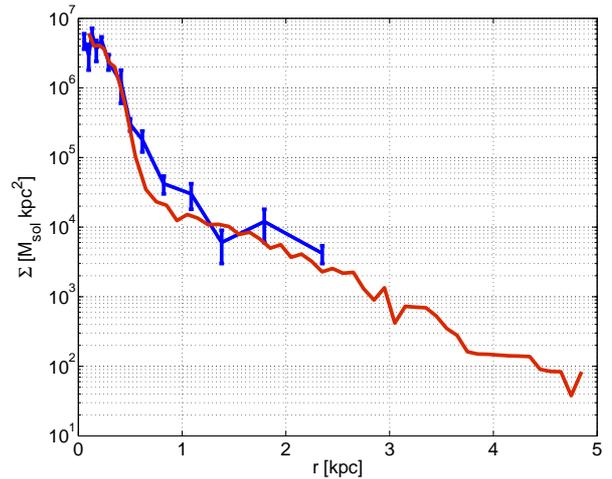}}
\caption{Surface density from observational data and the baryonic profile deduced from the observation. The model is in red and the data with the error bars are shown in blue.}
\label{SurfDen}
\end{figure}

Fig. \ref{SurfDen} shows a plot of the density profile and the data from \citet{2006ApJ...649..201M} where we have exploited the true physical dimensions instead of the angular dimensions for the system and the data representation because are an immediate indicator of the dynamical range we are considering (e.g. the surface on the celestial sphere can be calculated in spherical coordinates $\Sigma  = d_{\rm{Car}} \int_\alpha ^\beta  {d\varphi } \int_\alpha ^\gamma  {d\theta \sin \theta } $,
 where  $\alpha ,\beta ,\gamma $  are angular borders of the CCD mosaic of observational data, suitably converted from arcmin to radian.)
 
 An immediately evident result of our simulations is the reproduction the observed change in the slope of the surface density profile (e.g. \cite{2006ApJ...649..201M}) which fit the observational data quite well. 

\subsubsection{Radial velocity dispersion}
\begin{figure}
\resizebox{\hsize}{!}{\includegraphics{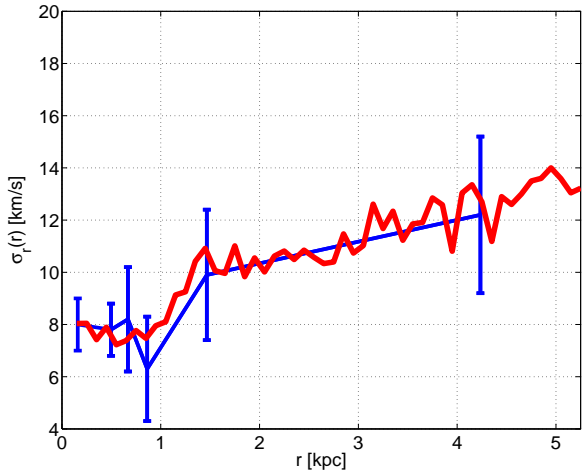}}
\caption{Radial velocity dispersion profile as computed from the model (in red) and the observation (in blue with the error bars). }
\label{DispVel}
\end{figure}

The other property we can now match is the radial velocity dispersion $\sigma _r \left( r \right)$ which is typically considered a fundamental indicator of the dynamical model of the galaxy if analyzed with the Jeans equation for spherical-stationary systems (Fig. \ref{DispVel}). The best-matching model nevertheless shows compatibility with the general trend of the baryonic matter traced by the observations.
The general decline of the velocity dispersion $1.5 kpc$ is also reproduced as a feature of the model (compare with the change of the slope of the surface density profile shown in Fig. \ref{SurfDen}. The innermost velocity dispersion value is around 8 ${\rm{km s}}^{ - 1}$.
 
\subsubsection{Star formation history}
At the last point we present the star formation history obtained from the N-body simulations. The criteria for star formation and the evolution of the isolated model have been adopted from \cite{2010A&A...514A..47P}. 
The resulting SFH for the orbiting model has been normalized here to the the highest observational peak in the star formation history deduced from the observations by \citet{2003ApJ...589L..85R}.

 This approach is mainly due to the impossibility of adequately resolving a dwarf galaxy with a one-to-one correspondence between stars and particles. The true galaxy Carina will contain probably several million of stars and only a small fraction of them can be represented with an N-body simulation. As a consequence,  $\frac{{dM}}{{dt}} $ depends on the resolution of our dwarf galaxy model. The consequence of this limitation is that the initial amount of gas that we have to assign to the dwarf galaxy model for Carina when it starts to evolve is still a free parameter that depends on the gas consumption that led to the formation of its oldest stellar population and on the gas consumption of the self-consistently triggered star formation.  Our goal has been to adapt the initial amount of gas such that there is no gas left after the last burst of star formation. This should be able to reproduce a Carina dwarf at the present time where essentially no gas is observed. 
 
 \begin{figure}
\resizebox{\hsize}{!}{\includegraphics{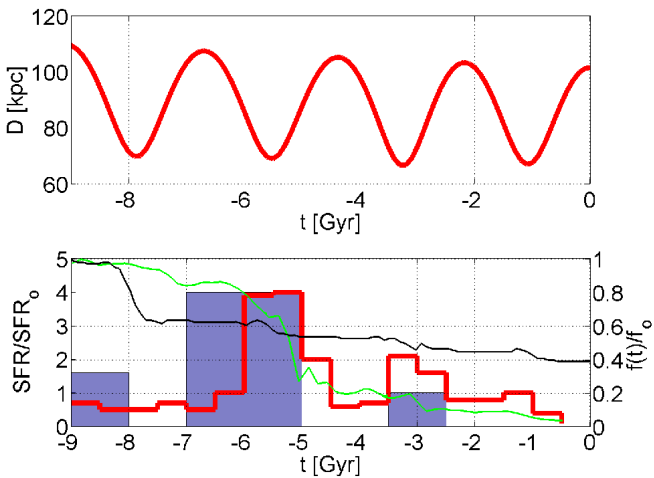}}
\caption{Star Formation Rate (SFR) as function of time normalized to the maximum of the observed SFR ($SFR_o$). The upper panel shows the radial distance of the N-body system to the origin of the reference system centered on the MW's barycentre for the best fitting orbit. The lower panel shows the star formation triggered by the pericentre passages and we reported the profile of SFR as deduced in \cite{2003ApJ...589L..85R}. In the lower panel green and black thin lines refer to the gas and bound mass fraction normalized to the initial value ($f\left( t \right)/f_0$ in the right vertical axis). }
\label{SFR}
\end{figure}

 While we are unable to impose strong constraints on this initial gas fraction and in order to reproduce the induced burst of star formation, we present in Fig. \ref{SFR} the simulated SFH normalized to the observed SFH. 
We emphasize that we are \textit{not} claiming to reproduce the true star formation history, but that we simply underline how tuned initial conditions are able to reproduce a system with present-day properties comparable with Carina including a star formation history dominated by well-separated star formation episodes. In particular, to obtain the star formation of the Carina dwarf galaxy we have to tune
\begin{itemize}
	\item a suitable amount of gas that survives the initial phase of SNII in the center of Carina and that initially coexists with the primordial oldest stellar population,
	\item  the original location of this dwarf galaxy that needs to be located not so far from the first pericentre passage. The first tidal pericentric passage may not extinguish the internal star formation processes.
	\item the amount of dark matter at the initial condition that will survive the first pericentre passage stripping. 
\end{itemize}
If the simulations are started prior to the apocentre passage ($t_{lb}<10$ Gyr), the first infall into the potential well of MW galaxy, which takes approximately 1 Gyr, causes too early a gas dissipation and dark matter stripping, making us unable to reproduce SFH-like Carina. For this reason, this alternative scenario of formation of the Carina dwarf galaxy has not been further investigated.

These conditions for the initial gas fraction and orbital initial conditions represent our chosen initial physical parameters to obtain the specific Carina SFH as an output that match the observations. As a general trend, we see from Fig. \ref{SFR} that the star formation after the first passage at the pericentre is typically more active. The galaxy produces an evident increase in the star formation rate around 6 Gyr ago and some smaller peaks correlated with the subsequent pericentre passages. 
A general trend is the evidence of the skewness of the star formation peaks, showing a rapid rise in the early phases and slower decline after the point of closest approach to the primary. In the same figure we present the evolution of the bound mass fraction (black line) and the gas fraction (green line). The initial amount of dark matter in the dwarf galaxy permits its baryonic component to make the first pericenter passage almost unperturbed, shielded within the dark matter potential well. The dark matter result is strongly stripped during the first pericenter passage, while the stripping results less important in the following pericenter passages. We find that gas consumption is roughly a monotonically decreasing function of time, controlled by the star formation episodes triggered by the orbital pericenter passages (see \cite{2009A&A...501..189R} for a different approach based on completely isolated models and different SNII recipes with respect to \cite{2010A&A...514A..47P}).

\section{Discussion and model prediction}\label{Conclusions}

In this paper, we presented an extensive treatment of the evolution of an orbiting Carina-like dwarf galaxy and we self-consistently derived the star formation history. 

Efforts have been made to consider all of the available observational constraints. Our methodology based on the minimum action, and developed in order to investigate all possible phase-space initial conditions, leads to a family of orbits able to reproduce with full N-body simulation a system compatible with the observational constraints. Once the orbital parameters that minimize the action in a realistic MW potential were derived, the dynamical properties and the star formation history of Carina were recovered. 

A previous study that attempted to interpret the observational constraints on Carina was recently presented by \citet{2008ApJ...679..346M}. However, in this paper the authors neglected the influence of star formation processes in their match with the observational data. Their work is based on a single component model, i.e., probably, the dark matter, which is tuned to match the kinematic properties of the baryonic matter, thus fitting directly a one-component model to match the observed properties. Another important difference between our work and the work by \citet{2008ApJ...679..346M} is in their use of a static external potential: by increasing the size of the MW halo during the time range of evolution (and keeping all the other parameter fixed) we can vary the instants of the pericentre passages of Carina dwarf galaxy in dependence on the initial conditions allowed by the proper motion error bars. 

In our work, the star formation history seems mainly driven by tidal processes, contrary to suggestions in other recent works (e.g., \citet{2003AJ....126.2346P}). To quantify the influence of the environment on the star formation rate is always a difficult task not only for dwarf galaxies (e.g., \citet{1994ApJ...435...22K}). Recently also \citet{2009ApJ...692.1305L} showed how in the local environment a bursty SFH has to be statistically preferred (see also, e.g., \citet{1997RvMA...10...29G, 1998ARA&A..36..435M, 2003ApJ...589L..85R}). Nevertheless, we do not claim to have found the true reason for the episodic star formation history that Carina experienced in the past. We just presented a possible interpretation in the framework of the dynamical interactions. The details of the interplay with the environment still need to be fully understood and reproduced for interacting dwarfs (e.g., \citet{2009AJ....137.3038B}). 
However, by tuning the amount of gas that is collapsing in a dark matter dominated protocloud of Carina (and that is retained despite initial phase of Supernovae type II explosions), a quite similar star formation history for the Carina can be reproduced self-consistently as result of a direct reaction of the system to the MW tidal field.

Once other processes will emerge as clearly influencing the amount of gas during the history of the Carina dwarf galaxy, e.g., the effects of ram-pressure stripping as possible key ingredient to clean the bulk of interstellar matter from dwarf galaxies (\citet{2003AJ....125.1926G}), then the initial amount of gas has to be tuned again, however without influence on our orbit determination. Nevertheless the expected density for the MW gaseous halo is probably too low (e.g., \citet{2000ApJ...529L..81M}) in order to strongly influence the gas evolution of the inner part of the Carina dwarf galaxy which orbits between 60 and 100 kpc from the MW center (see, e.g., Fig \ref{FigOrbits03}). 

Finally, the probability of an interaction with the Magellanic Clouds is considered in Appendix \ref{MCcol}.

\begin{acknowledgements}
We are grateful to an anonymous referee for improving this paper. Simulations were performed partially on the GRACE supercomputer (grants I/80 041-043 of the Volkswagen Foundation and 823.219-439/30 and /36 of the Ministry of Science, Research and the Arts of Baden-W\"urttemberg) and partially with the supercomputers at John von Neumann - Institut fuer Computing (NIC) (NIC-project number 2979). PB acknowledges the special support by the NAS Ukraine under the Main Astronomical Observatory GRAPE/GRID computing cluster project. PB's studies are also partially supported by the program Cosmomicrophysics of NAS Ukraine.

\end{acknowledgements}

\bibliographystyle{apj}
\bibliography{BiblioArt}

\appendix
\section{Determination of spatial velocities}
We update the transformation matrix ${\bm{T}}$ defined in \citet{1987AJ.....93..864J} (following \citet{1985spas.book.....G}) to equinox J2000:
	\[
\bm{T}=\left( {\begin{array}{*{20}c}
   { - 0.0534} & { - 0.8750} & { - 0.4810}  \\
   { + 0.4939} & { - 0.4418} & { + 0.7488}  \\
   { - 0.8678} & { - 0.1975} & { + 0.4558}  \\

 \end{array} } \right)
\]
The matrix ${\bm{B}}$ of \citet{1987AJ.....93..864J} remains defined as:
	\[
{\bm{B}} \equiv {\bm{T}} \cdot \left( {\begin{array}{*{20}c}
   {\cos (\alpha )\cos (\delta )} & { - \sin (\alpha )} & { - \cos (\alpha )\sin (\delta )}  \\
   {\sin (\alpha )\cos (\delta )} & {\cos (\alpha )} & { - \sin (\alpha )\sin (\delta )}  \\
   {\sin (\delta )} & 0 & {\cos (\delta )}  \\

 \end{array} } \right),
\]
where $\left( {\alpha ,\delta } \right)$
 are the equatorial coordinates of the generic dwarf galaxy at the present time $t=t_0 $. In our case we get  $\left( {\alpha ,\delta } \right)_{\rm{Car}}  = \left( {1.76, - 0.89} \right)rad$ for the Carina dwarf galaxy. Now we have to take into account a reflection of the velocity axis that we want pointing away from the Galactic center and we get the correction for the rotation of the galaxy as well as the motion of the sun relative to the local standard of rest:
	\[
\begin{gathered}
  \left( {\begin{array}{*{20}c}
   {v_x }  \\
   {v_y }  \\
   {v_z }  \\

 \end{array} } \right) = \left( {\left( {\begin{array}{*{20}c}
   { - 1} & 0 & 0  \\
   0 & 1 & 0  \\
   0 & 0 & 1  \\

 \end{array} } \right) \cdot \left( {\left( {\left. {{\bm{B}}\left( {\begin{array}{*{20}c}
   {v_r }  \\
   {k\mu _\alpha  d}  \\
   {k\mu _\delta  d}  \\

 \end{array} } \right)} \right)} \right.} \right.} \right. +  \\ 
   + \left. {\left. {\left( {\begin{array}{*{20}c}
   {v_{x \odot ,LSR} }  \\
   {v_{y \odot ,LSR} }  \\
   {v_{z \odot ,LSR} }  \\

 \end{array} } \right)} \right)} \right) + \left( {\begin{array}{*{20}c}
   0  \\
   {V_c }  \\
   0  \\

 \end{array} } \right) \\ 
\end{gathered} 
\]
where ${\bm{v}} = \left\{ {v_x ,v_y ,v_z } \right\}$
 is the velocity of the dwarf galaxy at the instant  $t = t_0 $, $v_r $
 is the radial velocity, $\left( {\mu _\alpha  ,\mu _\delta  } \right)$
 the proper motions in $\rm{\rm{arcsec}  \cdot \rm{yr}^{ - 1}} $, $d$ the distances that have to be assumed in pc to apply the conversion values $k = 4.74$; ${\bm{v}}_{ \odot ,LSR}  = \left\{ {10.0,5.2,7.2} \right\}  \rm{km \cdot s^{ - 1}} $ as in \citet{1998gaas.book.....B} and $V_c  = 220 \rm{km \cdot s^{ - 1}} $
 as adopted by the IAU (1986). 
For the distances, we also adopt a different reference system from \citet{1987AJ.....93..864J} in order to obtain a  right-handed reference system. This is achieved by imposing a double reflection for the positive X-axis originating from the Galactic center and the positive Y-axis pointing along decreasing Galactic longitude:
	\[
\left( {\begin{array}{*{20}c}
   x  \\
   y  \\
   z  \\

 \end{array} } \right) = \left( {\begin{array}{*{20}c}
   { - 1} & 0 & 0  \\
   0 & { - 1} & 0  \\
   0 & 0 & 1  \\

 \end{array} } \right).\left( {d\left( {\begin{array}{*{20}c}
   {\cos (b)\cos (l)}  \\
   {\cos (b)\sin (l)}  \\
   {\sin (b)}  \\

 \end{array} } \right)} \right) + \left( {\begin{array}{*{20}c}
   {x_ \odot  }  \\
   {y_ \odot  }  \\
   {z_ \odot  }  \\

 \end{array} } \right),
\]
where ${\bm{x}} = \left( {x,y,z} \right)$
 is the generic position of the dwarf galaxy, ${\bm{x}}_ \odot   = \left\{ {x_ \odot  ,y_ \odot  ,z_ \odot  } \right\} \cong \left\{ {8.5,0.0,0.0} \right\}$
 is the sun's position assumed for simplicity to lie in the plane of the MW and $\left( {l,b} \right)$ are the Galactic coordinates that can be consistently derived for a cross check from 
	\[
\left( {\begin{array}{*{20}c}
   {\cos (b)\cos (l)}  \\
   {\cos (b)\sin (l)}  \\
   {\sin (b)}  \\

 \end{array} } \right) = {\bm{T}}^{ - 1} \left( {\begin{array}{*{20}c}
   {\cos \left( \delta  \right)\cos \left( \alpha  \right)}  \\
   {\cos \left( \delta  \right)\sin \left( \alpha  \right)}  \\
   {\sin \left( \delta  \right)}  \\

 \end{array} } \right).
\]

The selection of an aligned reference system between the configuration space and the velocity space will permit us to treat, in a simpler form, the velocity vector as a derivative of the position vector for times different from the present. Similarly, we can derive the errors as 
	\[
\begin{gathered}
  \left( {\begin{array}{*{20}c}
   {\sigma _x^2 }  \\
   {\sigma _y^2 }  \\
   {\sigma _z^2 }  \\

 \end{array} } \right) = 2\mu _\alpha  \mu _\delta  \frac{{\sigma _p^2 }}
{{p^4 }}\left( {\begin{array}{*{20}c}
   {B_{12} B_{13} }  \\
   {B_{22} B_{23} }  \\
   {B_{32} B_{33} }  \\

 \end{array} } \right)k^2  +  \hfill \\
   + {\bm{B}}^2 \left( {\begin{array}{*{20}c}
   {\sigma _{v_r }^2 }  \\
   {\left( {\frac{k}
{p}} \right)^2 \left( {\left( {\frac{{\mu _\alpha  \sigma _p }}
{p}} \right)^2  + \sigma _{\mu _\alpha  }^2 } \right)}  \\
   {\left( {\frac{k}
{p}} \right)^2 \left( {\left( {\frac{{\mu _\delta  \sigma _p }}
{p}} \right)^2  + \sigma _{\mu _\delta  }^2 } \right)}  \\

 \end{array} } \right) \hfill \\ 
\end{gathered} 
\]
where $p$ is the parallax and $\bm{B}^2$ the matrix with elements $\left( {B_{ij} } \right)^2 $ $\forall i,j$.
With these equations and the proper motion and errors in the text, we can derive our estimate value for the orbital energy of the dwarf galaxy we are analyzing.

\section{Barycentre determination}\label{ApBar}
Once we try to move from the crude point mass determination to the full N-body description and include many different astrophysical aspects, we encounter the necessity to analyze the system properties of the many N-body system realizations by detecting the position of the center of mass of the system. The determination of the not inertial center of mass of a galaxy moving within an external force field can be performed in different ways but it is clear that we must evaluate it automatically to speed up the analysis of the large number of simulations we performed for finding the best match with the observational constraints. 
In this appendix, we present an original approach based on the downhill simplex method (Nelder \& Mead 1965). The downhill simplex method is a direct search method that works moderately well in low-dimensional stochastic problems. Our task is to apply the method to find the barycentre of the nucleus of our dwarf galaxy, limiting ourselves to the knowledge of the position of the galaxy at a given snapshot ${\bm{x}}_i \left( t \right) \in \mathbb{R}^3 $ for every particle, $\forall i$, and at every moment, $\forall t$, during the evolution of our dwarf. We refer the reader to books such as Numerical Recipes by \citet{1986nras.book.....P} which used the method in the section dedicated to the minimization already back in 1986. Here we limit ourselves to showing a further possible application where this method is suitable. In the practical implementation, we refer to the matrix formalism developed in the original work of Nelder and Mead. The simplex, a convex hull of a tetrahedron for our 3D space surrounding our MW galaxy be represented by a $3 \times \left( {3 + 1} \right)$
 time-dependent matrix, whose columns are the vertex 
 \[
 \alpha \left( {\hat t} \right) = \left( {\begin{array}{*{20}c}
   {B\left( {\hat t} \right)} & {{\bm{x}}_4 }  \\

 \end{array} \left( {\hat t} \right)} \right),
 \]
where $B \equiv \left( {{\bm{x}}_1 \left( {\hat t} \right),{\bm{x}}_2 \left( {\hat t} \right),{\bm{x}}_3 \left( {\hat t} \right)} \right)$. For any simplex $\alpha  \in \mathbb{R}^3 $
 we can define the matrix, $M = M\left( {\hat t} \right)$
 as the $3 \times 3$ matrix whose columns represent the edges of $\alpha \left( {\hat t} \right)$:
	\[
M\left( {\hat t} \right) = \left. {\left( {{\bm{x}}_1  - {\bm{x}}_4 ,{\bm{x}}_2  - {\bm{x}}_4 ,{\bm{x}}_3  - {\bm{x}}_4 } \right)} \right|_{t = \hat t}  = B\left( {\hat t} \right) - {\bm{x}}_4 {\bm{\hat e}}^T 
\]
and ${\bm{\hat e}} = \left( {1,1,1} \right)^T $. In this way we can, as a first step, check the degenerate character of the simplex by ensuring that the 3D volume of $\alpha \left( {\hat t} \right)$, $vol\left( {\alpha \left( {\hat t} \right)} \right) = \frac{1}{6}\left| {\det \left( {M\left( {\hat t} \right)} \right)} \right| > 0$. As a consequence, in a Euclidean geometry the reflection, expansion, inside/outside-contraction and shrinkage computed by the algorithm will always produce a non-degenerate tetrahedron. We define then the diameter of the simplex as $\emptyset \left( \alpha  \right) = \mathop {\max }\limits_{i \ne j} \left\| {{\bm{x}}_i  - {\bm{x}}_j } \right\|$ $\forall t$
 where $\left\| . \right\|$ is the standard norm. Finally we define a suitable function for finding the best point representing a barycentre of the dwarf galaxy. If we call $d_i  = \left\| {{\bm{x}}^{[i]}  - {\bm{x}}_b } \right\|$ the distance of the $i^{th} $-star from the guess value of the barycentre ${\bm{x}}_b $ at the instant $t = \hat t$ under consideration, then we need to maximize the function
\begin{equation}\label{eqApp01}
f\left( {\bm{x}} \right) = \sum\limits_{d_i  < r_* }^{} {m_i },
\end{equation}
where $r_ *  $ is a characteristic radius of the system we want to analyze. Our experience shows that it does not have to be physically related to the system, but a suitable choice of a few kpc can much improve the convergence velocity and the stability of the barycentre value if related to the convergence criteria. We know, in fact, that if $f$ is a bounded function then for every non-degenerate case it can be proved for the downhill simplex algorithm (we indicate with $k$ the iteration of the algorithm) that:
\begin{itemize}
	\item the sequence $\left\{ {f\left( {{\bm{x}}_1^{\left( k \right)} } \right)} \right\}$  \textit{always} converges;
	\item at every non-shrinking iterations $k$, $f\left( {{\bm{x}}_i^{\left( {k + 1} \right)} } \right) \leqslant f\left( {{\bm{x}}_i^{\left( k \right)} } \right)$
 for $1 \leqslant i \leqslant n + 1$, with strict inequality for at least one variable of I;
 \item if there is only a finite number of shrink iterations, then
 \begin{itemize}
 \item each sequence $\left\{ {f\left( {{\bm{x}}_i^{\left( k \right)} } \right)} \right\}$ converges as $k \to \infty $ for $1 \leqslant i \leqslant n + 1$;
 \item if we call $f\left( {{\bm{x}}_i^{\left( \infty  \right)} } \right) = \mathop {\lim }\limits_{k \to \infty } f\left( {{\bm{x}}_i^{\left( k \right)} } \right)$  then $f\left( {{\bm{x}}_i^{\left( \infty  \right)} } \right) \leqslant f\left( {{\bm{x}}_i^{\left( k \right)} } \right)\forall k$ and $1 \leqslant i \leqslant n + 1$;
 \item $f\left( {{\bm{x}}_1^{\left( \infty  \right)} } \right) \leqslant f\left( {{\bm{x}}_2^{\left( \infty  \right)} } \right) \leqslant ... \leqslant f\left( {{\bm{x}}_{n + 1}^{\left( \infty  \right)} } \right)$.
 \end{itemize}
 \item if there is only a finite number of non-shrink iterations, then all simplex vertexes converge to a single point.
\end{itemize}
All this ensures that the down hill simplex method applied to the function \eqref{eqApp01} will always converge if the number of shrink iteration is small (which is the case in our experience). We adopted a variable diameter as indicated by the subset of the stars for which we compute the barycentre, $\emptyset \left( \alpha  \right) < r_* :\mathop {\lim }\limits_{j \to \infty } r_*^{[j]} \left( {\hat t} \right) = 0$, e.g. with a physical radius converging to zero as the number of successive times that the down-hill simplex method is applied, $j$, increases the assigned$\forall t = \hat t$. In practice we of course chose $r_* $  to be slightly higher than the softening length of the gravitational potential $r_*  > \varepsilon _{star} $.

\section{Encounter with the Magellanic Clouds}\label{MCcol}
In this appendix we discuss the probability for a fly-by interaction with the Magellanic Clouds to investigate the possibility suggested by \citet{2006ApJ...649..201M}. We relegate these arguments to this appendix because they simply concern probability considerations that are derived from dynamics. The suggestion presented in the paper of \citet{2006ApJ...649..201M} comes from the analysis of only 15 stars in the local universe, too small to justify a further set of full simulations. 
In particular, we are interested in the recent determination of surprisingly large proper motion for the Large Magellanic Cloud (LMC) and the Small Magellanic Cloud (SMC) (\citet{2006ApJ...638..772K, 2006ApJ...652.1213K}), which caused  interest due to the possible implications for their evolutionary history (e.g. \citet{2009IAUS..256...93K, 2008AJ....135.1024P, 2007ApJ...668..949B, 2007ApJ...656L..61O}). As explained in \citet{2008arXiv0809.4263K}, the most relevant problem faced by the integration of the orbits of the Magellanic Clouds with an expected apocentre so far from the MW barycentre, is the unknown mass distribution of the MW for distances larger than 100-200 kpc (e.g. \citet{2007ApJ...668..949B}). 
With our MW galactic potential model presented in Appendix B, we cannot extend the integration of the LMC orbit much further as required to overlap the entire time with the orbit integration spanned by the minimization of the action for Carina. Thus we will limit ourselves to the last 3 Gyr of evolution in look-back time because in the potential of MW, the LMC after 3 Gyr of look-back time integration will already be more distant than 200 kpc from the MW barycentre, falsifying any phase-space derivation. Thus we will accept the conclusion already presented in \citet{2007ApJ...668..949B} in favor of a single/first pericentre passage for the MC not more than few hundred million years ago and we proceed by minimizing the distance function $d_{Car - MC} :\mathbb{R}^7  \to \mathbb{R}^ +  $

\[
d_{Car - MC}  = d_{Car - MC} \left( {{\bm{v}}_{0,Car} ,{\bm{v}}_{0,MC} ,t} \right)
\]
in the 7-dimensional space of the initial values for $
{\bm{v}}_{0,Car}  \in [{\bm{v}}_{0,Car}  \pm \delta {\bm{v}}_{0,Car} ]$, ${\bm{v}}_{0,MC}  \in [{\bm{v}}_{0,MC}  \pm \delta {\bm{v}}_{0,MC} ]$
 computed as in Appendix A  and $t \in \left] { - 3,0} \right]$ Gyr. We integrated the equation of motion here for the Magellanic Clouds, taking into account an extra term due to the dynamical friction as ascribed in Eqn. \eqref{eqsiF} caused by the dependence of the force on the square of the mass of the Magellanic Clouds (e.g. $m_{LMC}  \cong 2. \times 10^9 \rm{M_ \odot}  $) that makes the dynamical friction much more relevant than in the case of Carina.
 The result is that the closest distance approach permitted between the two galaxies in the range of the possible phase-space observational errors is more than 50 kpc, completely ruling out any possible interaction within the currently suggested phase-space error range deduced from the observations. The same consideration holds for an interaction with the SMC.

\end{document}